\documentclass{emulateapj}
\newcommand{\lsim}{\raisebox{0.3mm}{\em $\, <$} \hspace{-3.3mm}
\raisebox{-1.8mm}{\em $\sim \,$}}
\newcommand{\gsim}{\raisebox{0.3mm}{\em $\, >$} \hspace{-3.3mm}
\raisebox{-1.8mm}{\em $\sim \,$}}
\newcommand{\bm}[1]{\mbox{\boldmath $#1$}}

\shorttitle{FEEDBACK FROM SUPERCRITICAL DISK ACCRETION FLOWS}
\shortauthors{OHSUGA}

\begin{document}


\title{Feedback from supercritical disk accretion flows;
Two-dimensional radiation-hydrodynamic simulations
of stable and unstable disks with radiatively driven outflows}


\author{K. Ohsuga\altaffilmark{1}}
\affil{Department of Physics, Rikkyo University, 
Toshimaku, Tokyo 171-8501, Japan}

%\author{S. Djorgovski\altaffilmark{1,2,3} and Ivan R. King\altaffilmark{1}}
%\affil{Astronomy Department, University of California,
%    Berkeley, CA 94720}

%\author{C. D. Biemesderfer\altaffilmark{4,5}}
%\affil{National Optical Astronomy Observatories, Tucson, AZ 85719}
%\email{aastex-help@aas.org}

%\and

%\author{R. J. Hanisch\altaffilmark{5}}
%\affil{Space Telescope Science Institute, Baltimore, MD 21218}


%\altaffiltext{1}{Visiting Astronomer, Cerro Tololo Inter-American Observatory.
%CTIO is operated by AURA, Inc.\ under contract to the National Science
%\altaffiltext{2}{Society of Fellows, Harvard University.}
%\altaffiltext{3}{present address: Center for Astrophysics,
%    60 Garden Street, Cambridge, MA 02138}
%\altaffiltext{4}{Visiting Programmer, Space Telescope Science Institute}
%\altaffiltext{5}{Patron, Alonso's Bar and Grill}


\begin{abstract}
The supercritical disk accretion flow with 
radiatively driven outflows
is studied based on two-dimensional radiation-hydrodynamic simulations
for a wide range of the mass input rate,
$\dot{M}_{\rm input}$,
which is the mass supplied from the outer region
to the disk per unit time. 
The $\alpha$-prescription is adopted for the viscosity.
We employ $\alpha=0.5$, as well as $\alpha=0.1$
for $\dot{M}_{\rm input}\ge 3\times 10^2L_{\rm E}/c^2$
and only $\alpha=0.5$ 
for $\dot{M}_{\rm input}\le 10^2L_{\rm E}/c^2$,
where $L_{\rm E}$ is the Eddington luminosity 
and $c$ is the speed of light.
The quasi-steady disk and radiately driven outflows form
in the case in which the mass input rate
highly exceeds the critical rate,
$\dot{M}_{\rm input}>3\times 10^2 L_{\rm E}/c^2$.
Then, the disk luminosity as well as the kinetic energy output 
rate by the outflow exceeds the Eddington luminosity.
The moderately supercritical disk,
$\dot{M}_{\rm input}\sim 10-10^2 L_{\rm E}/c^2$,
exhibits limit-cycle oscillations.
The disk luminosity goes up and down across the Eddington luminosity,
and the radiatively driven outflows intermittently appear.
The time averaged mass, momentum, 
and kinetic energy output rates
by the outflow as well as the disk luminosity 
increase with increase of the mass input rate, 
$\propto \dot{M}_{\rm input}^{0.7-1.0}$
for $\alpha=0.5$
and 
$\propto \dot{M}_{\rm input}^{0.4-0.6}$
for $\alpha=0.1$.
Our numerical simulations show that
the radiatively driven outflow model 
for the correlation between black hole mass 
and bulge velocity dispersion
proposed by \citeauthor{SR98} and \citeauthor{King03}
is successful
if $\dot{M}_{\rm input}c^2/L_{\rm E} \sim$
a few 10 ($\alpha=0.5$) or $\gsim$ a few ($\alpha=0.1$).
\end{abstract}


\keywords{accretion: accretion disks --- black hole physics ---
galaxies: nuclei --- hydrodynamics --- radiative transfer}



\section{Introduction}
The evolution of supermassive black holes (BHs) and their host galaxies
is a major topic of current interest. 
The observations of high-redshift quasars showed
that the supermassive BHs had formed 
in the early universe
%when the universe was less than a gigayear old
\citep{Becker01,Fan01,Shields06}.
It has been reported that 
the gas accretion is a significant process
for evolution of the supermassive BHs \citep{YT02}.
If the supercritical accretion 
%(the mass accretion rate exceeds the Eddington rate, $L_{\rm E}/c^2$, 
(the mass accretion rate exceeds the critical rate, $L_{\rm E}/c^2$, 
where $L_{\rm E}$ is the Eddington luminosity and 
$c$ is the speed of light) is possible,
a growth timescale of BHs can be much shorter than the
Eddington timescale.
Thus, the supercritical accretion might resolve the
problem of the formation of supermassive BHs.

The supercritical accretion onto the central BH 
is also thought to play
important roles for the evolution of their host galaxies.
%When the luminosity of the active galactic nucleus (AGN) 
%becomes large enough, it drives the remaining gas from the
%nucleus of the host galaxy, ending black hole growth 
%as well as star formation 
%(Silk \& Rees 1998; Di Matteo et al. 2005, and references therein).
It has been suggested that the correlation between 
the velocity dispersion of the bulge stars ($\sigma_\star$)
and the BH mass 
\citep[$M_{\rm BH}-\sigma_\star$ relation;][]{Gebhardt00,FM00,GH06},
which implies that there is some physical link between
the evolution of the BHs and the host galaxies,
is established by the feedback from the supercritical accretion
flows.

\citet{King03} suggested that
the strong outflow from the supercritical accretion flow
regulates the evolution of the supermassive BH and its host galaxy,
leading to the $M_{\rm BH}-\sigma_\star$ relation 
\citep[see also][]{SR98}.
%King (2003) suggested that
%the strong outflow from the super-Eddington accretion flows
%regulates the evolution of the SMBHs and host galaxies,
%leading the $M-\sigma$ relation (see also Silk \& Rees 1998).
%In their model, the interstellar matter is blown away 
%from the host galaxy, since a large amount of energy and momentum 
%is supplied by the outflow from the super-Eddington accretion flow
%around the central black hole.
%Then, the fuelling onto the black hole stops and
%the $M-\sigma$ relation is naturally established.
However, the momentum and kinetic energy output rates 
by the outflow
were assumed without being carefully treated,
even though they are significant physical quantities in this scenario.
%since the outflow produced by the super-Eddington accretion flow
%leads to $M_{\rm BH}-\sigma_\star$ relation.

\citet{Umemura01} suggested that the $M_{\rm BH}-\sigma_\star$ relation
is built up by mass accretion onto the galactic center
via the radiation drag \citep[see also][]{KU02}.
%In their model, the interstellar matter 
%loses angular momentum via radiation drag by the 
%stellar radiation in the galactic bulge and falls onto the
%galactic center.
Since mass accretion is caused 
by the stellar radiation of the bulge stars,
the BH mass is correlated with the bulge mass.
Though the mass accretion rate exceeds the critical rate
in this mechanism,
the feedback from the supercritical accretion 
is not taken into consideration.

Although the standard disk model,
which was proposed by \citet{SS73},
elucidates the fundamental properties of 
luminous compact objects,
it breaks down in the 
supercritical accretion regime.
In this case,
a large part of the photons generated inside the disk via the viscous process
is advected inward and swallowed by the BH with 
accreting matter.
This is the so-called photon trapping in the disk accretion flows
\citep{O02,OMW03,SM03}.
The radiation pressure becomes dominant over the gas pressure
and drives the radiatively driven outflows.
The circular motion as well as the convection occurs in the disk.
They are basically multi-dimensional effects.
These effects are not correctly treated in the slim disk 
\citep{ACLS88},
%which is widely believed to describe the supercritical accretion flows,
since it is a radially one-dimensional model.

The two-dimensional radiation hydrodynamic (RHD) simulations 
of the supercritical disk accretion
flows around the BHs were initiated by \citet{ECK88}
and improved by \citet{Okuda02}.
\citet{O05} for the first time succeeded in 
reproducing the quasi-steady structure
of the supercritical disk accretion flows.
\citet{O06} performed the numerical simulations
of the unstable disks 
with the moderately supercritical accretion rate.
However, 
%only a few cases of mass accretion rate was investigated and 
the outflow from the supercritical accretion flow 
is poorly understood.
Whereas the radiatively driven outflow has extensively been studied
by many researches \citep[e.g.,][]{BB77,WF99},
they treat the disk as an external radiation source
and do not solve the radiation transfer.

In this paper, 
by performing two-dimensional RHD simulations
of stable and unstable disks,
%we report the outflows of the super-Eddington accretion flows
%for the wide range of the mass input rate,
we report the feedback (mass, momentum, and energy output rates)
from the supercritical disk accretion flows around BHs
for a wide range of the mass input rate,
which is the mass supplied 
from the outer region to the disk per unit time.
%We also investigate the quasi-steady disk accretion flows
%whose mass input rate highly exceeds the critical rate
%and the limit-cycle behaviour of the unstable disks
%with moderately supercritical rate.

\section{BASIC EQUATIONS  AND NUMERICAL METHOD}
Basic equations and our numerical methods are described 
in detail in \citet{O05} and \citet{O06}.
The set of RHD equations including the viscosity term
is solved by 
an explicit-implicit finite difference scheme on the Eulerian grids.
Here, we use spherical polar coordinates $(r, \theta, \varphi)$,
where $r$ is the radial distance, 
$\theta$ is the polar angle,
and $\varphi$ is the azimuthal angle.
The origin is set at a central BH.
Since we assume axisymmetry (i.e., $\partial/\partial \varphi=0$),
as well as reflection symmetry
relative to the equatorial plane (with $\theta = \pi/2$),
the computational domain can be restricted to one quadrant
of the meridional plane.
The domain consists of spherical shells of 
$3r_{\rm S} \leq r \leq 500r_{\rm S}$ 
and $0 \leq \theta \leq \pi/2$,
and is divided into $96\times 96$ grid cells,
where $r_{\rm S}$ is the Schwarzschild radius.
We describe the gravitational field of the BH
in terms of pseudo-Newtonian hydrodynamics,
in which the gravitational potential
is given by $-GM_{\rm BH}/(r-r_{\rm S})$ as was 
introduced by \citet{PW80}.
The flow is assumed to be non-self-gravitating.
The basic equations are the continuity equation,
\begin{equation}
  \frac{\partial \rho}{\partial t}
%  + {\rm div}(\rho {\bm v}) = 0,
  + \nabla \cdot (\rho {\bm v}) = 0,
  \label{mass_con}
\end{equation}
the equations of motion,
%\begin{equation}
\begin{eqnarray}
  \frac{\partial (\rho v_r)}{\partial t}
%  + {\rm div}(\rho v_r {\bm v}) 
  &+& \nabla \cdot (\rho v_r {\bm v}) 
  = - \frac{\partial p}{\partial r} \\
  &+& \rho \left[ 
    \frac{v_\theta^2}{r} + \frac{v_\varphi^2}{r}
    -\frac{GM_{\rm BH}}{(r-r_{\rm S})^2}
  \right]
  + f_r
%  + \rho \frac{\kappa+\sigma}{c} F_r
  + q_r,
  \label{mom_r}
\end{eqnarray}
%\end{equation}
\begin{equation}
  \frac{\partial (\rho r v_\theta)}{\partial t}
%  + {\rm div}(\rho r v_\theta {\bm v}) 
  + \nabla \cdot (\rho r v_\theta {\bm v}) 
  = - \frac{\partial p}{\partial \theta}
  + \rho v_\varphi^2 \cot\theta
  + r f_\theta
  + r q_\theta,
  \label{mom_th}
\end{equation}
\begin{equation}
  \frac{\partial (\rho r v_\varphi \sin\theta)}{\partial t}
%  + {\rm div}(\rho r \sin\theta\cdot v_\varphi {\bm v}) 
  + \nabla \cdot (\rho r v_\varphi \sin\theta {\bm v}) 
  = q_\varphi r \sin\theta ,
  \label{mom_varphi}
\end{equation}
the energy equation of the gas,
\begin{equation}
  \frac{\partial e }{\partial t}
%  + {\rm div}(e {\bm v}) 
  + \nabla\cdot(e {\bm v}) 
%  = -p\, {\rm div}{\bm v} -4\pi \kappa B 
  = -p\nabla\cdot{\bm v} -4\pi \kappa B 
  + c\kappa  E_0 + \Phi_{\rm vis},
  \label{gase}
\end{equation}
and the energy equation of the radiation,
\begin{equation}
  \frac{\partial E_0}{\partial t}
%  + {\rm div}(E_0 {\bm v}) 
  + \nabla\cdot(E_0 {\bm v}) 
%  = -{\rm div}{\bm F_0} -\nabla{\bm v}:{\bm {\rm P}_0}
  = -\nabla\cdot{\bm F_0} -\nabla{\bm v}:{\bm {\rm P}_0}
  + 4\pi \kappa B - c\kappa E_0,
  \label{rade}
\end{equation}
where $\rho$ is the mass density,
%Here, $\rho$ is the density,
$\bm{v}=(v_r, v_\theta, v_\varphi)$ is the velocity,
$p$ is the gas pressure,
%$M$ is the mass of the central object (BH or NS),
$e$ is the internal energy density of the gas,
$B$ is the blackbody intensity,
$E_0$ is the radiation energy density,
${\bm F}_0$ is the radiative flux,
${\bm {\rm P}}_0$ is the radiation pressure tensor,
%$f_r$ and $f_\theta$ are the radial and polar component 
%of the radiation force, 
$\kappa$ is the absorption opacity,
${\bm q}=(q_r, q_\theta, q_\varphi)$ is the viscous force,
and $\Phi_{\rm vis}$ is the viscous dissipative function.

The radiation force ${\bm f}_{\rm rad} = (f_r, f_\theta)$ 
is given by 
\begin{equation}
  {\bm f}_{\rm rad} = \frac {\chi}{c} {\bm F}_0,
  \label{frad}
\end{equation}
where $\chi (= \kappa+\rho \sigma_{\rm T}/m_{\rm p})$ 
is the total opacity 
with $\sigma_{\rm T}$ being the Thomson scattering cross-section
and $m_{\rm p}$ being the proton mass.
For the absorption opacity,
we consider the free-free absorption 
and the bound-free absorption for solar metallicity
\citep{HHS62,RL79}.
%We consider the radiation pressure force on 
%the Thomson scattering, the free-free absorption, and 
%the bound-free absorption for solar metallicity
%(Hayashi, Hoshi, \& Sugimito 1962, Rybicki \& Lightman 1979).
In the present study, 
the outflow is mainly accelerated 
by the radiation force on the Thomson scattering,
although the radiation force on the free-free and bound-free absorption is 
also taken into consideration.

As the equation of state, we use
\begin{equation}
  p=(\gamma-1)e,
\end{equation}
%where $\gamma$ is the specific heat ratio and is set to the value of $5/3$.
where $\gamma$ is the specific heat ratio.
The temperature of the gas, $T$, can then be calculated from
\begin{equation}
  p=\frac{\rho k T}{\mu m_{\rm p}},
\end{equation}
where $k_{\rm B}$ is the Boltzmann constant and
$\mu$ is the mean molecular weight.
%We set $\mu=0.5$ in the present study.

The radiative flux and the radiation stress tensor are
solved under the flux limited diffusion
approximation \citep{LP81},
so that they are expressed 
in terms of the radiation energy density
\citep{TS01}.
It gives correct relations in the optically thick diffusion limit
and optically thin streaming limit, respectively.

In this paper, we assume that 
only the $r\varphi$-component of the viscous stress tensor
%is non zero.
is nonzero, leading $q_r=q_\theta=0$.
This component plays important roles for
the transport of the angular momentum
and heating of the disk plasma.
In our form, 
it is proportional to $\alpha p_{\rm total}$
in the optically thick limit
and $\alpha p$
in the optically thin limit,
where $\alpha$ and $p_{\rm total}$ are
the viscosity parameter and the total pressure.
Our viscosity model is basically 
the same as the $\alpha$ prescription of the viscosity
proposed by \citet{SS73}.

We start the calculations with a hot, rarefied, and 
optically thin atmosphere. 
There is no cold dense disk initially.
We assume that matter
is continuously injected into the computational domain
through the outer disk boundary 
($r=500r_{\rm S}$, $0.45\pi \leq \theta \leq 0.5\pi$).
%Therefore, 
%we can avoid the influence of the initial configuration on the final result,
%though a long integration time is required.
The injected matter is supposed to 
have a specific angular momentum corresponding to the 
Keplerian angular momentum at $r=100r_{\rm S}$.
We set the injected mass accretion rate 
(mass input rate, $\dot{M}_{\rm input}$)
so as to be constant at the boundary.

%Throughout the present study, 
%we assume the BH mass to be $M_{\rm BH}=10M_\sun$.
Throughout the present study, 
we assume $M_{\rm BH}=10M_\sun$, $\gamma=5/3$, and $\mu=0.5$.
The $\alpha$ parameter is set to be $0.5$ and $0.1$.
%but we represent the results only in the case of $\alpha=0.5$
%for the moderately supercritical regime,
%$\dot{M}_{\rm input}=10-10^2 L_{\rm E}/c^2$.
%(We discuss about 
%$\alpha$ dependence of our results in \S 3.3.)
%We set the viscosity parameter to be 
%$\alpha=0.1$ for 
%$\dot{M}_{\rm input} \geq 3\times 10^2 L_{\rm E}/c^2$
%and $\alpha=0.5$ for 
%$\dot{M}_{\rm input} \leq 10^2 L_{\rm E}/c^2$.
The time step is $\sim 3\times 10^{-6}$ s, which is determined by
$\min\left[ \Delta r/c, r\Delta\theta/c \right]$,
where $\Delta r$ and $\Delta\theta$ are 
the cell sizes in the radial and polar directions, respectively.

\section{RESULTS}
%\subsection{Time evolution}
\subsection{Highly supercritical case}
\begin{figure}[b]
\epsscale{1.1}
%\plotone{stableA05.ps}
\plotone{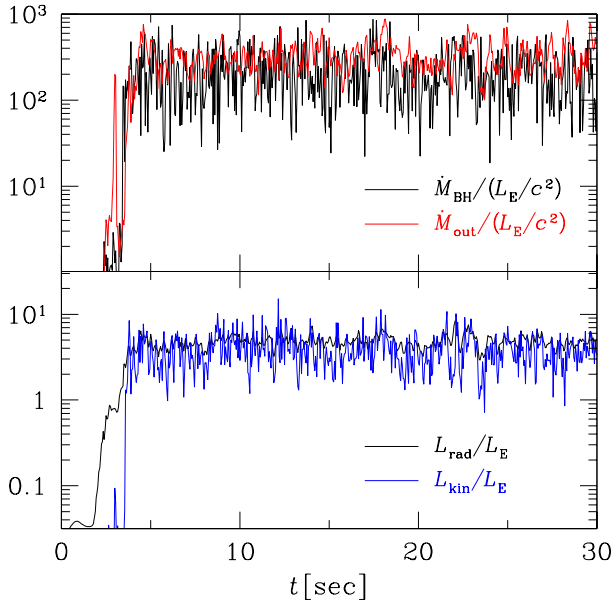}
 \caption{
 Time evolution of the mass accretion rate onto the BH 
 ({\it black line in the top panel}), 
 the mass output rate ({\it red line}), 
 the luminosity ({\it black line in the bottom panel}), 
 and kinetic energy output rate ({\it blue line})
 for $\dot{M}_{\rm input}=10^3 L_{\rm E}/c^2$.
 The viscosity parameter $\alpha$ is set to be $0.5$.
 \label{stable}
 }
\end{figure}

In Figure \ref{stable}
we plot the time evolution 
of the mass accretion rate onto the BH $\dot{M}_{\rm BH}$ 
({\it black line in the top panel}),
the luminosity $L_{\rm rad}$ ({\it black line in the bottom panel}),
and the mass and kinetic energy output rates
$\dot{M}_{\rm out}$ ({\it red line}) 
and $L_{\rm kin}$ ({\it blue line})
for $\dot{M}_{\rm input}=10^3 L_{\rm E}/c^2$,
where the viscosity parameter is set to be $0.5$.
The luminosity is evaluated
by integrating the radiative flux at the outer boundary.
It almost equals the luminosity of the disk.
The mass and kinetic energy output rates
indicate the mass and kinetic energy 
ejected through the outer boundary per unit time 
by the outflow of $v_r>v_{\rm esc}$,
where $v_{\rm esc}$ is the escape velocity.
%\begin{equation}
% \dot{M}_{\rm out}= \int 2\pi r^2 \rho(r,\theta) v_r(r,\theta)
%  \sin\theta d\theta 
%  \,\,\,\,\,\mbox{if $v_r(r,\theta)>v_{\rm esc}(r,\theta)$},
%\end{equation}
%and
%\begin{equation}
% L_{\rm out}= \int 2\pi r^2 \frac{1}{2}\rho(r,\theta) v_r^3(r,\theta)
%  \sin\theta d\theta 
%  \,\,\,\,\,\mbox{if $v_r(r,\theta)>v_{\rm esc}(r,\theta)$}.
%\end{equation}
%\newpage
%Fig. 1

Since the matter injected from the outer disk boundary
has angular momentum, 
it accumulates in the computational domain
without accreting onto the BH
%in the early evolutional phase ($t\lsim 7$ sec).
in the early evolutional phase ($t\lsim 3$ s).
%The matter starts to accrete onto the BH at $t\sim 7$ sec.
The matter starts to accrete onto the BH at $t\sim 3$ s.
This critical time roughly coincides with the viscous
timescale (see equation [2] in \citeauthor{O05} [\citeyear{O05}]).
%The mass accretion leads to the outflow
%which is accelerated by the strong radiation pressure.
%As shown in this figure, 
All the quantities stay nearly constant
%when $t\gsim 7$ s.
when $t\gsim 3$ s.
The total mass contained within the computational domain
is also constant in this phase.
%Here, it is noted that that mass input rate almost equals
%the sum of the mass accretion rate 
%and total mass ejected from the outer boundary per unit time 
%in the final phase, $t\gg 7$s.
Thus, we conclude that the flow is quasi steady in this phase.
The radiation pressure-dominated thick disk forms,
and the radiatively driven outflows appear above and below the disk.
%Detailed time evolution and quasi-steady structure 
%are shown in Ohsuga et al. (2005).
The quasi-steady structure is similar to that in \citet{O05},
although they employ $\alpha=0.1$.

As shown in Figure \ref{stable},
the mass accretion rate 
is about 2 orders of magnitude larger than the critical rate,
%and the luminosity stays around $3 L_{\rm E}$
and the luminosity stays around $4.8 L_{\rm E}$
in the quasi-steady state.
We also find that the strong and quasi-steady outflow is generated. 
%in the case that the mass input rate highly exceeds the critical rate.
The mass output rate is comparable to the mass accretion rate.
The kinetic energy output rate
exceeds the Eddington luminosity,
whereas it is slightly smaller than the luminosity.
The energy conversion efficiency, 
$\left(L_{\rm rad}+L_{\rm kin}\right)/\dot{M}_{\rm BH}c^2$,
is much smaller than the prediction of the standard disk, $\sim 0.1$.
This is because a large amount of photons generated inside the disk
are swallowed by the BH without being radiated away by the photon trapping.

%Here, we note that our numerical simulations reveals that
%the quasi-steady disk and strong outflow forms 
%in the case of $\dot{M}_{\rm input}=3\times 10^2$, 
%$3\times 10^3$, 

\subsection{Moderately supercritical case}
\begin{figure}[b]
 \epsscale{1.1}
 \plotone{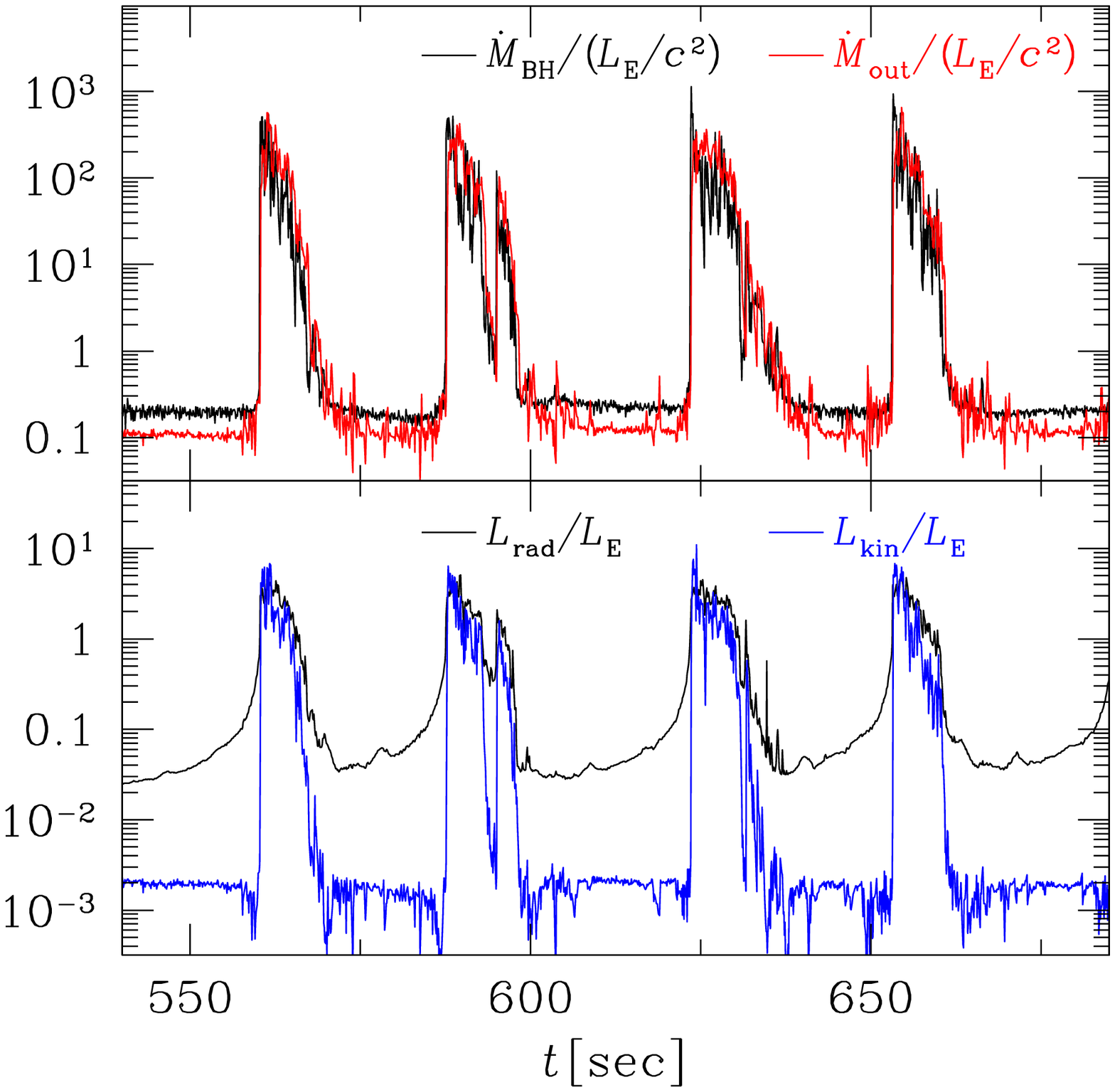}
 \caption{
 Same as Fig. \ref{stable}, 
 but for $\dot{M}_{\rm input}=10^2L_{\rm E}/c^2$.
 \label{unstable100}
 }
\end{figure}

Next, we show the time evolution in the case in which
the mass input rate moderately exceeds the critical rate
in Figure \ref{unstable100}.
Here, we set the mass input rate to be $10^2L_{\rm E}/c^2$
and the viscosity parameter to be $0.5$.
In this case, the mass accretion rate
drastically varies, although the matter is continuously injected
into the computational domain at constant rate.
The mass accretion rate suddenly rises
from $\sim 0.3 L_{\rm E}/c^2$ to $\sim 300 L_{\rm E}/c^2$,
and it decays after 10 s.
Such time variation of the mass accretion rate
occurs at intervals of $\sim 60$ s
and triggers off the luminosity oscillations.
The luminosity is around $0.03L_{\rm E}$ 
in the low-luminosity state 
and $3L_{\rm E}$ in the high-luminosity state.
The photon trapping reduces the luminosity in the high state.

%Fig. 2

Such disk oscillations are induced by the
disk instability in the radiation pressure-dominated region.
The disk theory predicts that
the radiation pressure-dominated disk is unstable
if the radiative cooling is dominant over the advective cooling
\citep{LE74,SH75,KFM98}.
%Pringle 1976; Shakura \& Sunyaev 1976).
Since the advective cooling becomes the main cooling process
when the mass accretion rate is much higher than the 
critical rate, the disk with a highly supercritical accretion rate 
is stabilised.
%as shown in the previous subsection.
In contrast, the moderately supercritical disk exhibits
%limit-cycle oscillations as shown in Figure \ref{unstable100}.
limit-cycle oscillations.

In Figure \ref{unstable100},
we also find that the strong outflow appears
only in the high-luminosity state.
In this state, the mass output rate
is about 2 orders of magnitude larger than the critical rate.
The kinetic energy output rate exceeds the Eddington luminosity
and is comparable to the luminosity.
In contrast, it is found that $L_{\rm kin}$ is much smaller 
than $L_{\rm rad}$ in the low-luminosity state.
It implies that the matter is not effectively accelerated 
by the radiation force
when the disk shines at sub-Eddington luminosity.
The radiatively driven outflows are intermittently produced
in the case in which
the mass input rate moderately exceeds the critical rate.

The mass ejected from
the computational domain per unit time
(sum of the mass accretion rate onto the BH 
and the mass ejection rate through the outer boundary)
almost equals the mass input rate on average.
Thus, simulated limit-cycle behaviour is not a transient phenomenon.
It would continue 
as long as the mass input rate does not change so much.
Here, we note that 
such a balance between mass input and ejection rates
is for the first time reproduced in the present simulations,
since our integration time is much longer than that 
in \citet{O06}
and the relatively large viscosity parameter is employed.

%although it has not been achieved in Ohsuga (2006).
%In the present study, the integration time is longer 
%than that in Ohsuga et al. (2006),
%and the relatively large viscosity parameter is employed ($\alpha=0.5$).

\begin{figure}
\epsscale{1.1}
%\epsscale{0.8}
\plotone{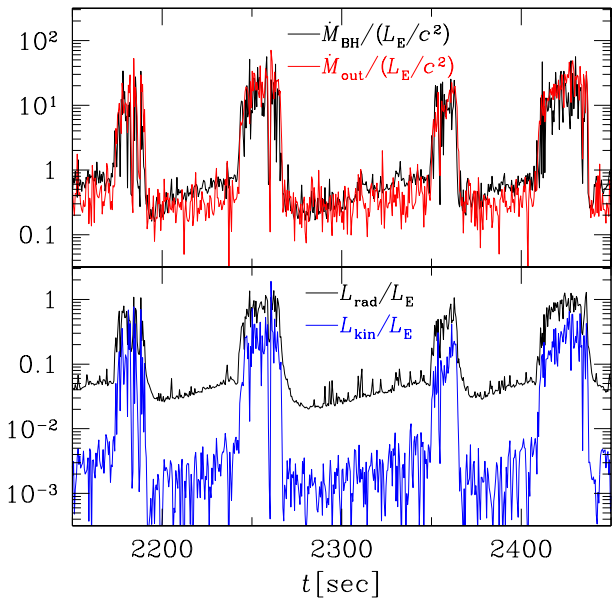}
%\plotone{unstable.eps}
\caption{
Same as Fig. \ref{stable},
 but for $\dot{M}_{\rm input}=10L_{\rm E}/c^2$.
\label{unstable10}
}
\end{figure}

Our simulations show that 
the disk is also unstable and 
exhibits the limit-cycle oscillations
in the case of $\dot{M}_{\rm input}=10 L_{\rm E}/c^2$
(see Fig. \ref{unstable10}).
%The resulting time evolution is similar to 
%that for $\dot{M}_{\rm input}=10^2 L_{\rm E}/c^2$.
%the burst occurs at about 100 sec intervals.
%However, as compared with the case of 
As compared with the case of
$\dot{M}_{\rm input}=10^2 L_{\rm E}/c^2$,
the bursting behaviour is not remarkable.
The luminosity,
as well as the kinetic energy output rate,
is comparable to or less than the Eddington luminosity
even in the high-luminosity state.

%Fig. 3

%Here, we need to remark that 
%a very long integration time is required 
%to investigate the limit-cycle oscillations 
%with using the small viscosity parameter,
%although we employ relatively large viscosity parameter
%in the present simulations.
%%though 
%%we employ relatively large viscosity parameter $\alpha=0.5$.
%%This is because 
%%the viscous timescale is roughly in proportion to 
%%$\alpha^{-1}$,
%%although the time step of the simulations is determined by
%%the speed of light (see \S2).
%More importantly, we need radiation magneto-hydrodynamic (MHD) simulations,
%since the source of disk viscosity is likely to be of magnetic origins
%(Hawley, Balbus, \& Stone 2001; Machida, Matsumoto, \& Mineshige 2001)

%\subsection{Radiatively driven outflow}
\subsection{Feedback}
\begin{figure}
% \epsscale{.80}
 \epsscale{1.1}
% \plotone{fbA05_2.ps}
 \plotone{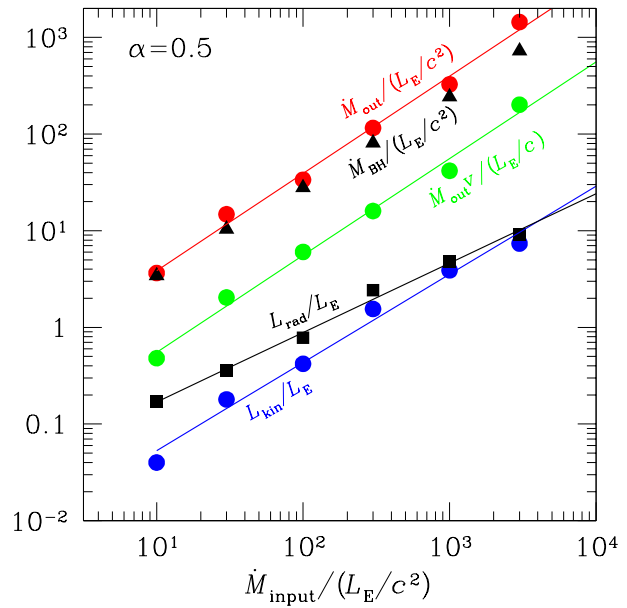}
 \caption{
 Time-averaged mass output rate ({\it red circles}),
 the kinetic energy output rate ({\it blue circles}),
 and the momentum output rate ({\it green circle}) by the outflow
 for $\alpha=0.5$.
 The mass accretion rate onto the BH
 and the luminosity are plotted with triangles and squares.
 Red, blue, green, and black lines represent the fitting
 formulae of equation (\ref{outA05}-\ref{radA05}).
 \label{fb}
 }
\end{figure}

Based on the results of our numerical simulations for $\alpha=0.5$,
we represent in Figure \ref{fb}
the time averaged 
mass accretion rate onto the BH ({\it triangles}),
luminosity ({\it squares}),
mass output rate ({\it red circles}),
kinetic energy output rate ({\it blue circles}),
and momentum output rate $\dot{M}_{\rm out}v$ ({\it green circles})
as functions of the mass input rate.
Here, the momentum output rate
%means the momentum in the radial direction
means the radial component of the momentum
carried out of the computational domain per unit time
through the outer boundary
by the outflow of $v_r>v_{\rm esc}$.
The disks with $\dot{M}_{\rm input}>3\times 10^2L_{\rm E}/c^2$
are stable.
In contrast, 
the disks exhibit the limit-cycle behaviour 
%in the case of $\dot{M}_{\rm input}=10$, $30$, and $10^2$.
in the case of $\dot{M}_{\rm input}=10-10^2L_{\rm E}/c^2$.
%In this case, the strong outflow is generated quasi periodically.

%Fig. 4

We find that all quantities monotonously 
rise with an increase of $\dot{M}_{\rm input}$.
%They gently rise in the highly supercritical regime, 
%$\dot{M}_{\rm input}c^2/L_{\rm E} \gg 10^2$.
%In contrast, the mass, kinetic energy,
%and momentum output rates suddenly decrease
%in the near-critical regime, $\dot{M}_{\rm input}c^2/L_{\rm E}=10$.
%in contrast with $L_{\rm rad}$ and $\dot{M}_{\rm BH}$.
%It implies that the matter is not effectively 
%accelerated by the radiation pressure.
%The near-critical disks can not produce the strong 
%radiatively driven outflows.
%If the mass input rate is less than
%$\dot{M}_{\rm input} \lsim L_{\rm E}/c^2$,
%the standard disk which is not accompanied by the outflow 
%would form.
%If the mass input rate is less than the critical rate,
%the standard disk which is not accompanied by the outflow 
%would form.
%
%Here, we present simple fitting formulae for our numerical results;
%feedback from the supercritical disk accretion flows,
We present simple fitting formulae 
with the least squares method for our numerical results;
feedback from the supercritical disk accretion flows 
in the case of $\alpha=0.5$ are given by
\begin{equation}
% \log\left(\frac{\dot{M}_{\rm out}}{L_{\rm E}/c^2}\right)
%  =0.76x+0.28+\frac{\log(0.3x)}{x},
 \log\left(\frac{\dot{M}_{\rm out}}{L_{\rm E}/c^2}\right)
  =\log\left(\frac{\dot{M}_{\rm input}c^2}{L_{\rm E}}\right)-0.42,
  \label{outA05}
\end{equation}
\begin{equation}
% \log\left(\frac{L_{\rm kin}}{L_{\rm E}}\right)=0.61x-1.3+\frac{\log(0.3x)}{x},
 \log\left(\frac{L_{\rm kin}}{L_{\rm E}}\right)
  =0.91\log\left(\frac{\dot{M}_{\rm input}c^2}{L_{\rm E}}\right)-2.2,
  \label{kinA05}
\end{equation}
\begin{equation}
% \log\left(\frac{\dot{M}_{\rm out}v}{L_{\rm E}/c}\right)
%  =0.76x-0.56+\frac{\log(0.3x)}{x},
 \log\left(\frac{\dot{M}_{\rm out}v}{L_{\rm E}/c}\right)
  =\log\left(\frac{\dot{M}_{\rm input}c^2}{L_{\rm E}}\right)-1.3,
  \label{momA05}
\end{equation}
\begin{equation}
% \log\left(\frac{L_{\rm rad}}{L_{\rm E}}\right)=0.66x-1.35,
 \log\left(\frac{L_{\rm rad}}{L_{\rm E}}\right)
  =0.72\log\left(\frac{\dot{M}_{\rm input}c^2}{L_{\rm E}}\right)-1.5.
  \label{radA05}
\end{equation}
These fitting formulae are also plotted 
by red, blue, green, and black lines in Figure \ref{fb},
respectively.
Here, it is noted that they are valid only 
in the supercritical accretion regime.
%If the mass input rate is less than the critical rate,
If the mass input rate is comparable to or less than the critical rate,
the standard disk, which is not accompanied by the outflow, would form.
%unless the radiation force on lines accelerates the outflow
%(discussed later).

\begin{figure}
% \epsscale{.80}
 \epsscale{1.1}
% \plotone{fbA01.ps}
 \plotone{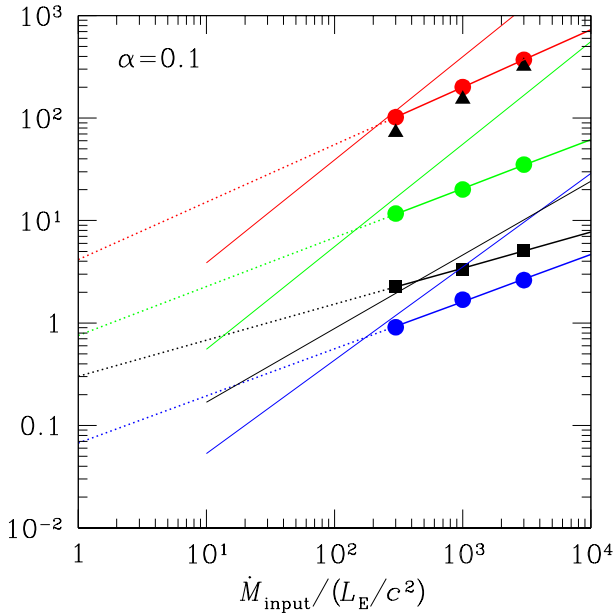}
 \caption{
 Same as Fig. \ref{fb}, but for $\alpha=0.1$.
 Thick solid lines represent the fitting formulae of equation 
(\ref{outA01}-\ref{radA01}).
 They are extended onto the small mass input regime ({\it dotted lines}).
 The fitting formulae for $\alpha=0.5$
 are plotted by thin lines.
 \label{fbA01}
 }
\end{figure}

Whereas Figure \ref{fb} shows the feedback effects 
in the case of $\alpha=0.5$ as functions of the mass input rate,
they are sensitive to the viscosity parameter.
In Figure \ref{fbA01}
we show the resulting feedback effects for $\alpha=0.1$.
Here, there are no results for 
$\dot{M}_{\rm input}<10^2L_{\rm E}/c^2$.
We need to consume much time in the numerical simulations
for a small mass input rate because of the long viscous timescale.
In this figure, thick solid lines represent the fitting formulae
for $\alpha=0.1$.
They are 
\begin{equation}
 \log\left(\frac{\dot{M}_{\rm out}}{L_{\rm E}/c^2}\right)
  =0.56\log\left(\frac{\dot{M}_{\rm input}c^2}{L_{\rm E}}\right)+0.62
  \label{outA01}
\end{equation}
\begin{equation}
 \log\left(\frac{L_{\rm kin}}{L_{\rm E}}\right)
  =0.46\log\left(\frac{\dot{M}_{\rm input}c^2}{L_{\rm E}}\right)-1.2,
  \label{kinA01}
\end{equation}
\begin{equation}
 \log\left(\frac{\dot{M}_{\rm out}v}{L_{\rm E}/c}\right)
  =0.48\log\left(\frac{\dot{M}_{\rm input}c^2}{L_{\rm E}}\right)-0.12,
  \label{momA01}
\end{equation}
and
\begin{equation}
 \log\left(\frac{L_{\rm rad}}{L_{\rm E}}\right)
  =0.35\log\left(\frac{\dot{M}_{\rm input}c^2}{L_{\rm E}}\right)-0.52.
  \label{radA01}
\end{equation}
Though they are derived based on the results for 
$\dot{M}_{\rm input}>3\times 10^2L_{\rm E}/c^2$,
we extend them onto the small mass input regime ({\it dotted lines}).
The fitting formulae for $\alpha=0.5$ are also plotted 
by thin lines.
As shown in this figure,
we find that feedback effects are more sensitive to the
mass input rate for $\alpha=0.5$ than for $\alpha=0.1$.
It is also found that the fitting formulae for the two cases
intersect around 
$\dot{M}_{\rm input}=3\times 10^2L_{\rm E}/c^2$.

%Fig. 5

Since \citet{King03} proposed that
the $M_{\rm BH}-\sigma_\star$ relation
is established via radiatively driven outflow 
from the supercritical accretion flow onto the supermassive BH 
in the galactic center.
The momentum output rate 
$\dot{M}_{\rm out}v$ 
is assumed to be $\sim L_{\rm E}/c$
in their model.
Thus, our results imply that their outflow model is successful
as long as the normalized mass input rate, 
$\dot{M}_{\rm input}c^2/L_{\rm E}$,
is a few tens in the case of $\alpha=0.5$
(see the green line in Fig. \ref{fb}).
In the case of $\alpha=0.1$,
the fitting formula for the momentum output rate 
predicts that the King model is successful 
for $\dot{M}_{\rm input}c^2/L_{\rm E} \sim$ a few
(see the green dotted line in Fig. \ref{fbA01}).
However, we should note that
the fitting formulae for $\alpha=0.1$ are derived 
based on the results for high mass input rate
and there are no numerical results for small mass input rate.
The fitting formulae might overestimate the power of 
the radiatively driven outflow around 
$\dot{M}_{\rm input} \sim L_{\rm E}/c^2$.
This is because 
the radiation force, which is comparable to the gravity,
cannot efficiently accelerate the outflow.
In this case,
the normalized mass input rate for establishment of
the $M_{\rm BH}-\sigma_\star$ relation
would shift towards the higher $\dot{M}_{\rm input}$ side.
We need further studies to understand the $\alpha$ dependence
in detail.

%%Although we set $\alpha$ to be $0.5$ in the prsent study,
%When we employ the relatively small viscosity parameter,
%the feedback effects are slightly weakened.
%We find $L_{\rm kin}$ and $L_{\rm rad}$ are
%smaller by factors of $2.8$ and $1.7$ for $\alpha=0.1$ than
%for $\alpha=0.5$ in the case of 
%$\dot{M}_{\rm input} = 3\times 10^3 L_{\rm E}/c^2$.
%We also find 
%that such discrepancy is attenuated as the mass input rate decreases.
%In the case of 
%$\dot{M}_{\rm input} = 3\times 10^2 L_{\rm E}/c^2$,
%$L_{\rm kin}$ is a factor of $1.5$ 
%smaller for $\alpha=0.1$ than for $\alpha=0.5$,
%and $L_{\rm rad}$ for the case $\alpha=0.1$ almost equals
%that for $\alpha=0.5$.
%[Here, we note that
%an extremely long integration time is required 
%to confirm the feedback from the unstable disks
%with small viscosity parameter.
%Even in the case of $\alpha=0.5$,
%we need to calculate up to $\sim 2500$ sec
%for $\dot{M}_{\rm input} = 10 L_{\rm E}/c^2$
%(see Fig. \ref{unstable10}).]
%Therefore, the feedback effects in the case of $\alpha=0.1$
%is relatively insensitive to the mass input rate
%as compared with the case of $\alpha=0.5$.

\section{DISCUSSION}
\subsection{Formation of supermassive black holes}
Our simulations show that the highly supercritical disk is stable
and the moderately supercritical disk exhibits 
limit-cycle oscillations.
%The formation scenario of the supermassive BHs
%is a major topic of current interest.
The physical mechanism for mass supply 
from the host galaxy onto the galactic center
is one of the most hotly debated issues.
%is necessary to form the supermassive BH.
Although its time evolution is not understood yet,
the normalized mass input rate 
($\propto \dot{M}_{\rm input}/M_{\rm BH}$)
would gradually decrease as the seed BH grows into 
the supermassive BH.
Thus, our results imply that the seed BH grows via the supercritical 
disk accretion flow in the early phase,
and subsequently, the BH accretion disk evolves into 
the sub-critical phase by way of the oscillation phase.
The growth timescale via supercritical accretion is
$M_{\rm BH}/\dot{M}_{\rm BH} = 4.5\times 10^6 (\dot{M}_{\rm BH}c^2/10^2L_{\rm E})^{-1} \rm yr$.

Here, we note that our results would not change that much 
even if we employ a large BH mass ($M_{\rm BH}\gg 10M_\odot$),
since the radiation force on the Thomson scattering 
mainly supports the disk and accelerates the outflow
in our simulations.
The gas density decreases with an increase of the BH mass, 
$\rho \propto M_{\rm BH}^{-1}$, 
on the condition that the normalized mass input rate is kept constant.
Hence, the radiation force on the Thomson scattering ($\propto \rho$)
is dominant over that on the bound-free and free-free absorptions
($\propto \rho^2$) 
even in the case of massive BHs.

\citet{Umemura01} suggested the model for quasar formation,
by which the mass is supplied onto the galactic center
via the radiation drag, forming the supermassive BH in $\sim 10^9$ yr.
Although 
the mass supply rate onto the galactic center 
highly exceeds the critical rate in their model,
they assumed the accretion rate onto the BH
to be the critical rate.
%The rapid growth of the BH due to the 
The supercritical accretion onto the central BH
would lead the emergence of quasars in the early evolutional phase
($\ll 10^9$ yr).
%The growth timescale of the BH via supercritical accretion is
%$M_{\rm BH}/\dot{M}_{\rm BH} = 4.5\times 10^6 (\dot{M}_{\rm acc}c^2/10^2L_{\rm E})^{-1} \rm yr$.
The radiation and radiatively driven outflow
would affect the stellar evolution of the host galaxy.
%through the change of the distribution and the physical condition
%of the inter stellar medium.
It is not taken into consideration in their model.
Especially, if the metallicity of the gas is much smaller 
than the local interstellar value,
it is known that the ultraviolet radiation 
plays important roles on the star formation activity
\citep{HRL97,SU06}.

\subsection{Outflows}
The velocity of the radiatively driven outflow in our simulations
is $0.1-0.3c$.
It is roughly consistent with the results from the observations 
of quasars.
The observations of the blueshifted X-ray absorption lines
have revealed 
that quasars eject the highly ionized matter with a velocity of 
$\sim 0.1c$
\citep{Pounds03a,Pounds03b,ROW03}.
%Note that the optical depth to the Thomson scattering 
%is independent of the black hole mass,
%though we employ $M_{\rm BH}=10M_\odot$ in the present simulations.

Line opacities are not 
taken into consideration in our simulations.
The outflow would be enhanced 
via the radiation force on lines,
unless the matter is highly ionized by 
strong radiation fields \citep{PSK00,CN03,PK04}.
The line force is thought to be one of the acceleration mechanisms
of the outflows in active galactic nuclei (AGNs).
The AGNs likely have powerful outflows 
even for $L_{\rm rad}<L_{\rm E}$.

\citet{Proga99} studied the disk winds driven by line force
for $L_{\rm rad}<L_{\rm E}$,
based on the stellar wind model given by \citet{CAK75}.
%The disk wind driven by the line force is studied 
%by \citet{Proga99} in the case of sub-Eddington luminosity,
%whereas we focus on the the case with $L_{\rm rad}\gsim L_{\rm E}$ 
%in the present study.
They concluded that the mass output rate scales with the luminosity
%with $L_{\rm rad}/L_{\rm E}$ 
as $\dot{M}_{\rm out} \propto (L_{\rm rad}/L_{\rm E})^{1-2.5}$.
%It is roughly consistent with our results,
Our results do not deviate from their result.
We derive the relations of
$\dot{M}_{\rm out} \propto (L_{\rm rad}/L_{\rm E})^{1.4}$
from equations (\ref{outA05}) and (\ref{radA05}),
as well as 
$\dot{M}_{\rm out} \propto (L_{\rm rad}/L_{\rm E})^{1.6}$
from equations (\ref{outA01}) and (\ref{radA01}).
%$\dot{M}_{\rm out} \propto (L_{\rm rad}/L_{\rm E})^{1.4-1.6}$
%$\dot{M}_{\rm out} \propto (L_{\rm rad}/L_{\rm E})^{1.4}$
%for $\alpha=0.5$ and
%$\dot{M}_{\rm out} \propto (L_{\rm rad}/L_{\rm E})^{1.6}$
%for $\alpha=0.1$
%[see equations (\ref{outA05}), (\ref{radA05}), 
%(\ref{outA01}) and (\ref{radA01})].
Note that the disk is treated as the source of radiation and mass
without solving its structure in their work.
In contrast, both the disk and the outflow are solved
in our simulations.
In addition, the central BH swallows a large amount of matter in our work.

The magnetic effects might also enhance the outflow.
It has been reported by magnetohydrodynamic (MHD) simulations that 
the magnetized accretion disks produce the outflows,
although these simulations focus on 
the radiatively inefficient accretion flow
\citep{HB02,KMS02,KMS04}.

%It is noted that the outflow would be enhanced
%via the radiation force on absorption lines 
%in the case of massive BHs (Proga, Stone, \& Kallman 2000;
%Chelouche \& Netzer 2003).
%Line opacities are not 
%taken into consideration in our simulations.
%The magnetic effects might also enhance the outflow.
%It has been reported by the magneto-hydrodynamic (MHD) simulations that 
%the magnetic outflow is generated (Hawley \& Balbus 2002;
%Kudoh, Matsumoto, \& Shibata 2002; Kato, Mineshige, \& Shibata 2004), 
%although these simulations focus on 
%the radiatively inefficient accretion flow.
%%where the mass accretion rate is much smaller than the
%%Eddington rate 

\subsection{Future work}
In this paper
we investigate the feedback from the 
supercritical disk accretion flows around BHs
as functions of the mass input rate.
%for the wide range of the mass input rate
Although we employ merely two viscosity parameters,
$\alpha=0.5$ and $0.1$,
the feedback effects might be more sensitive to the 
viscosity parameter than to the mass input rate.
In addition, we do not present results for 
$\dot{M}_{\rm input}<10^2L_{\rm E}/c^2$
in the case of $\alpha=0.1$.
Further long integration time is required 
to simulate the accretion flows with a small viscosity parameter 
and/or a small mass input rate because of the long viscous timescale. 
Such numerical simulations are left as future work.

More importantly, we need radiation MHD simulations,
%We stress again that radiation MHD simulations are need,
since the source of disk viscosity is likely to be of 
magnetic origin \citep{SPB99,SP01,HBS01,MMM01}.
A more physical treatment of the viscosity might lead 
to the difference from the $\alpha$-disks.
Whereas local radiation MHD simulations have been performed recently
\citep[e.g.,][]{Turner03},
global simulations should be explored in future work.

\section{CONCLUSIONS}
By performing the two-dimensional RHD simulations,
we study the feedback (mass, momentum, and energy output rates)
from the supercritical disk accretion flows around BHs
for the wide range of the mass input rate,
$\dot{M}_{\rm input}=10-3\times 10^3L_{\rm E}/c^2$.
%The adopted viscosity parameter, $\alpha$, is $0.5$ and $0.1$,
%but we present only results of $\alpha=0.5$ 
%for $\dot{M}_{\rm input}<10^2L_{\rm E}/c^2$.
The adopted viscosity parameter $\alpha$ is $0.5$ and $0.1$
for $\dot{M}_{\rm input}\ge 3\times 10^2L_{\rm E}/c^2$
and only $0.5$
for $\dot{M}_{\rm input}\le 10^2L_{\rm E}/c^2$.
The detailed study for $\alpha$-dependence 
is left as future work.
We summarize our results as follows.

\smallskip
1. 
In the case of high mass input rate,
$\dot{M}_{\rm input}>3\times 10^2 L_{\rm E}/c^2$,
the quasi-steady disk forms and the outflows driven by
the radiation force appear above and below the disk.
Both energy output rates by the radiation and the outflow
exceed the Eddington luminosity.

2.
The disk with moderately supercritical rate,
$\dot{M}_{\rm input}\sim 10-10^2 L_{\rm E}/c^2$,
exhibits limit-cycle oscillations,
leading to intermittent outflows.
The strong outflow is generated only in the high-luminosity state,
in which the energy output rates by the radiation and the outflow
are comparable to or exceed the Eddington luminosity.

3.
The time-averaged mass, momentum, 
and kinetic energy output rates
by the outflow, as well as the disk luminosity,
scale with mass input rate 
as $\propto \dot{M}_{\rm input}^{0.7-1.0}$
for $\alpha=0.5$.
In the case of $\alpha=0.1$, we find that 
they are proportional to $\dot{M}_{\rm input}^{0.4-0.6}$.

4.
Our numerical simulations show that
the radiatively driven outflow model 
for the $M_{\rm BH}-\sigma_\star$ relation
proposed by Silk \& Rees and King is successful,
as long as the normalized mass input rate
$\dot{M}_{\rm input} c^2/L_{\rm E}$
is around a few 10 ($\alpha=0.5$) or a few ($\alpha=0.1$).

%The detailed study of $\alpha$-dependence 
%is left as future work.

\acknowledgments

We would like to thank the anonymous reviewer 
for many helpful suggestions, which greatly improved
the original manuscript. 
%We especially thank K. Watarai, M. Mori, H. Susa, and N. Shibazaki
%for useful comments and discussions.
We especially thank J. P. Ostriker, S. Mineshige, and 
H. Susa for useful comments and discussions.
The calculations were carried out 
by a parallel computer at Rikkyo University
and Institute of Natural Science, Senshu University.
%
%This work is supported in part by 
%Research Fellowships of the Japan Society
%for the Promotion of Science for Young Scientists, 02796 (KO).
We acknowledge a research grant from the Japan Society 
for the Promotion of Science (17740111).
%The calculations were carried out at 
%Department of Physics in Rikkyo University.
%This work is supported in part by 
%Research Fellowships of the Japan Society
%for the Promotion of Science for Young Scientists, 02796 (KO),
%by the Grant-in-Aid of the Ministry of Education,
%Culture, Science, and Sports, 14740132, 
%by the Promotion and Mutual Aid Corporation for Private Schools of Japan
%(MM),
%by the Grants-in-Aid of the
%Ministry of Education, Science, Culture, and Sport, 
%(14079205, 16340057),
%and by a Grant-in-Aid for the 21st Century COE
% {\lq\lq}Center for Diversity and Universality in Physics" (SM).

\end{document}